\documentstyle[aps,12pt]{revtex}
\begin{document}
\tightenlines
\title{FRW cosmological model inside an isolated
Schwarzschild black hole}
\author{C. Ortiz$^1$, J.J. Rosales$^2$, J. Socorro$^1$ and  V. I. Tkach$^1$}
 \address{$^1$Instituto de F\'{\i}sica de la Universidad de Guanajuato,\\
 A.P. E-143, C.P. 37150, Le\'on, Guanajuato, M\'exico\\
$^2$Facultad de Ingenier\'{\i}a Mec\'anica El\'ectrica y
Electr\'onica, Universidad de Guanajuato\\
Prolongaci\'on Tampico 912, Bellavista, Salamanca, Guanajuato, M\'exico }
\maketitle
\begin{abstract}
Using the canonical quantum theory of spherically symmetric pure
gravitational systems, we present a direct correspondence between
the Friedmann-Robertson-Walker (FRW) cosmological model in the
interior of a Schwarzschild black hole and the nth energy
eigenstate of a linear harmonic oscillator. Such type of universe
has a quantized mass of the order of the Planck mass and harmonic
oscillator wave functions.
\end{abstract}


\section{Introduction}

 In the absence of a fundamental understanding of physics at very
 high energies and, in particular, in the absence of a consistent
 quantum theory of gravity, there is no hope, at present, to meet
 an understanding of the quantum origin of the Universe in a
 definitive way. However, in order to come nearer to this
 seemingly unattainable goal, it appears desirable to develop
 highly simplified, but consistent models, which contain as many
 as possible of those features which are believed to be present in a
 future complete quantum theory of gravity. It is believed that
 black holes may play an important role in our attempts to shed
 some light on the nature of a quantum theory of gravity.

Thus, in the framework of quantum gravity, black holes must be
treated as quantum objects. As such, they are characterized by
quantum numbers like mass, electric charge and angular momentum.
For neutral, non-rotating Schwarzschild black hole, the only
quantum number which is left is the mass $M$. Classically, it is
related to the area $A$ of the black hole horizon by the relation
$ A = \frac{16\pi G^2 M^2}{c^4}$, where $G$ is the Newtonian
gravitational constant. Important questions in black hole physics
are what the spectrum of $A$ looks like and what the degeneracies
of states are for a given values of $A$.
\\

The quantization of black holes was first proposed by Bekenstein
some years ago \cite{bekenstein}. The fundamental idea of his work
is the remarkable observation that the horizon area of a
non-extremal black hole behaves as a classical adiabatic
invariant. But in the spirit of Ehrenfest principle
\cite{Ehrenfest}, any classical adiabatic invariant should
correspond to a quantum entity with discrete spectrum. Bekenstein
conjectured that the horizon area of a quantum black hole should
have a discrete spectrum with uniformly spaced eigenvalues of the
form

\begin{equation}
A_n = \gamma l^2_{pl} n, \qquad\qquad n = 1,2,3,... \label{area}
\end{equation}
where $\gamma$ is a dimensionless constant to be determined, and
$l_{pl} = (\frac {G}{c^3})^{1/2}\hbar $ is the Planck length.
Bekenstein's proposal implies that the energy eigenvalues
corresponding to the stationary states of the black holes are

\begin{equation} \rm E_n = \sigma \sqrt{n} E_{pl}, \qquad
n=1,2,\cdots, \qquad E_{pl}= \sqrt{\frac{\hbar c^5}{G}},
\label{energy}
\end{equation}
where $\sigma = \sqrt{\frac{\gamma}{16\pi}}$ is of the order of
unity.

Using a combination of thermodynamics and statistical physics
arguments it was found in \cite{Bekenstein-Mukhanov} that the
dimensionless constant $\gamma$ should be of the form $\gamma =
\frac{4}{n} \ln\alpha $, where $\alpha$ is the degeneracy factor
of the $nth$ area level. Recently, Hod \cite{Hod} employed Bohr's
correspondence principle and found evidence in favor of the value
$\alpha = 3^n$.

On the other hand, Rosen \cite{rosen} put the equations of General
Relativity for the case of a closed homogeneous isotropic universe
in the form of a Schr\"odinger equation for the s-state of a
hidrogen-like atom, and was able to obtain the relation $\rm m_n=
\sqrt{n \pi}M_{pl}$ for the quantization of the mass spectrum.

The main goal of this work is to obtain a time independent
Schr\"odinger equation for the case of the closed FRW model inside
a black hole horizon. This is done following the canonical
quantization procedure by means of which, we show that our system
looks like a linear quantum oscillator. Besides, the area and
energy quantum spectra are obtained, as well as, the wave function
of the system.

\section{The canonical Hamiltonian}

We will start with the classical closed FRW universe inside a
Schwarzschild black hole horizon, which is the simplest model
which exhibits the collapse phenomenon. It must be filled either
with gravitational radiation or with some other form of energy.
Classically, the only black hole observable parameter for an
asymptotic external observer at rest is the black hole mass $M$.
It is well known, that this mass $M$ measured by this kind of
observer coincides with the mass $M$ in the usual expression of
the Schwarzschild metric. Then, the observer can define the
concept of energy $E$ of the black hole as $ E= Mc^2$
\cite{bryce,regge}, where $c$ is the velocity of light in vacuum.

To know the energy spectrum of the black hole, the observer should
derive a quantum equation for the system. In this work we will
consider a time-dependent metric inside a black hole horizon.

In the interior region, we write the line element as in \cite{ADM}
(using comoving hyperspherical coordinates ($\chi,\theta,\phi$)
for the star's interior and putting the origin of coordinates at
the star's center)

\begin{equation}
\rm ds^2= -N^2(t) c^2 dt^2 + R^2(t)\left[ d\chi^2 + sin^2 \chi
\left( d^2 \theta + sin^2 \theta d^2\phi \right) \right],
\label{metric}
\end{equation}
where $\rm N(t)$ and $\rm R(t)$ are the lapse function and the
scale factor, respectively.

Outside the black hole horizon the line element is
\begin{equation}
\rm ds^2 = -N^2(t)\left(1-\frac{R_s}{r} \right) c^2 dt^2
+\frac{1}{1-\frac{Rs}{r}} dr^2 + r^2\left( d^2 \theta + sin^2
\theta d^2\phi \right), \label{schw}
\end{equation}
where $\rm R_s=\frac{2MG}{c^2}$ is the Schwarzschild radius. The
connection between the two metrics (\ref{metric}) and (\ref{schw})
is made by the matching conditions on the star surface \cite{ADM}.

The spatial volume for the metric (\ref{metric}) is
\begin{equation}
\rm V =  R^3 \int_0^{\chi_0} d\chi \int_0^\pi d\theta
\int_0^{2\pi} d\phi sin^2 \,\chi sin\, \theta = R^3 \pi^2,
\end{equation}
where $\chi_0$ is determined by the matching conditions, which for
the case of black hole give $\chi_0=\frac{\pi}{2}$.

We consider a pure gravity system with Einstein action and the
corresponding term for the total energy \cite{bryce}

\begin{equation}
\rm S= \int \left[ -\frac{c^2}{2G N} R \dot R^2 + \frac{c^4}{2G} N R
-N E_s \right] dt,
\label{black}
\end{equation}
where $\rm E_s= M c^2$ is the ADM energy. Here, the last term in
the action (\ref{black}) corresponds to the surface integral at
spatial infinity, which leads to the Schwarzschild black hole mass
\cite{bryce}.

The action (\ref{black}) preserves the invariance under time
reparametrization

\begin{equation}
\delta t = a(t),
\label{repara}
\end{equation}
if the transformations of $\rm N(t)$ and $\rm R(t)$ are defined as
\begin{equation}
\rm \delta N = \frac{d(a N)}{dt} , \qquad \delta R = a \frac{d
R}{dt}. \label{transforma}
\end{equation}
Note that if we take the lapse function as
\begin{equation}
\rm N(t) = \tilde N(t) R(t) \frac{c^2}{M_{pl}G},
\label{nueva}
\end{equation}
and substituting into (\ref{black}), we have the following
invariant action
\begin{equation}
\rm S= \int \left[ -\frac{M_{pl}}{2\tilde N} \dot R^2 +
\frac{c^6}{2M_{pl}G^2} \tilde N R^2- \tilde N \frac{Mc^4}{M_{pl}G}
R \right] dt. \label{black1}
\end{equation}
Using the relations (\ref{transforma},\ref{nueva}), it is easy to show that
$\rm \tilde N(t)$
transforms as
\begin{equation}
\rm \delta \tilde N = \frac{d(a \tilde N)}{dt}\label{tilde}
\end{equation}

Proceeding with the Hamiltonian analysis we define the usual
canonical momentum conjugate to the $R(t)$ coordinate, $\rm
P_R=\frac{\partial L}{\partial \dot R}$ and performing the
Lagrange transformation, we can obtain the following canonical
Hamiltonian
\begin{eqnarray}
\rm H_{can} &=&\rm \tilde N \left[ -\frac{P_R^2}{2M_{pl}}
- \frac{c^6}{2M_{pl}G^2}
R^2 + \frac{M}{M_{pl}}\frac{c^4}{G} R   \right] \nonumber \\
&=& \rm
\tilde N \left[ -\frac{P_R^2}{2M_{pl}}-\frac{M_{pl}}{2} \omega^2_0
\left( R - \frac{MG}{c^2} \right)^2 +\frac{M}{2M_{pl}} Mc^2
\right].
\label{hamiltonian}
\end{eqnarray}
where $\rm \omega_0=\frac{c^3}{M_{pl}G}$ is the fundamental
frequency of the system. This form of the canonical Hamiltonian
explains the fact, that the lapse function $\rm \tilde N(t)$ is a
Lagrange multiplier, which enforces the only first class
constraint $\rm H=0$. The latter manifests the invariance of the
action under reparametrization transformations
(\ref{repara},\ref{transforma}). According to the Dirac's
constraint quantization procedure, the wave function must be
annihilated by the operator version of the classical constraint.

We transform Eq. (\ref{hamiltonian}) by defining $\rm \xi =R-
\frac{MG}{c^2}$, thus its momentum conjugate becomes $\rm P_\xi =
P_R$ and the constraint at the classical level reads as follows
\begin{equation}
\rm H_{can}= \tilde N H = \tilde N\left[ -\frac{P^2_\xi}{2M_{pl}}
-\frac{M_{pl}}{2} \omega^2_0 \xi^2  +\frac{M}{2M_{pl}} Mc^2 \right] =0,
\label{constraint}
\end{equation}
that can be also rewritten as
\begin{equation}
\rm \frac{P^2_\xi}{2M_{pl}} +\frac{M_{pl}\omega_0^2}{2} \xi^2 =
\frac{M}{M_{pl}} \frac{Mc^2}{2} = \frac{M}{M_{pl}}\frac{E_s}{2}.
\label{con}
\end{equation}

\section{Harmonic oscillator equation and quantization rules}
Making the usual realization of the operator $\rm \frac{\hat
P_\xi^2}{2M_{pl}} = - \frac{\hbar^2}{2M_{pl}} \frac{d^2}{d\xi^2}$
and applying it to the wave-function $\rm \psi$, we get the
following linear harmonic oscillator equation
\begin{equation}
\rm \left[- \frac{\hbar^2}{2M_{pl}}\frac{d^2}{d\xi^2} +
\frac{M_{pl}\omega_0^2}{2} \xi^2  \right] \psi= \frac{M}{M_{pl}}
\frac{E_s}{2} \psi. \label{wave}
\end{equation}
We can obtain the following relations $\rm E_s= \frac{c^4}{2G}
R_s$, $\rm E_{pl}=\frac{c^4}{G} \ell_{pl}$ (this relation is
equivalent to that appearing in (\ref{energy})), and making the
transformation $\frac{\xi}{\ell_{pl}}= x$, one can rewrite
(\ref{wave}) as
\begin{equation}
\rm \frac{1}{2} \left[ x^2 - \frac{d^2}{dx^2}\right] \psi=
\frac{1}{4}\frac{R_s  E_s}{\ell_{pl} E_{pl}} \psi. \label{wave1}
\end{equation}
Using the creation-annihilation representation,
\begin{eqnarray}
\rm a&=& \frac{1}{\sqrt{2}} \left( x + \frac{d}{dx}\right), \\
\rm a^{\dag} &=& \frac{1}{\sqrt{2}} \left(x - \frac{d}{dx}\right),
\end{eqnarray}
with the usual algebra between them, $[a,a^{\dag}]=1$, we can
rewrite Eq. (\ref{wave1}) as
\begin{equation}
\rm a^{\dag} a \psi = \frac{1}{2} \left[ x^2 -
\frac{d^2}{dx^2}\right] \psi -\frac{1}{2}\psi= \left( -\frac{1}{2}
+ \frac{1}{4} \frac{R_s  E_s}{\ell_{pl} E_{pl}} \right) \psi =  n
\psi, \qquad n=0,1,2,\cdots.
\end{equation}
In this way, we have the following useful relations
\begin{eqnarray}
R_s  E_s &=& 4(n+ \frac{1}{2}) \ell_{pl} E_{pl} \nonumber\\
&=& 4(n+ \frac{1}{2}) \hbar c, \label{uno} \\
E_s^2 &=& 2(n+\frac{1}{2}) E_{pl}^2, \label{energyn}\\
\frac{E_s}{2}&=&(n+\frac{1}{2}) \hbar \omega_0 \label{dos}.
\end{eqnarray}
It is clear that the system, even in its lowest energy state $\rm
n=0$, has a finite, minimal energy. Eq. (\ref{uno}) implies the
following quantization mass rule
\begin{equation}
\rm M_n= \sqrt{2n +1} \,\,M_{pl}, \label{rule mass}
\end{equation}
and Eq. (\ref{energyn}) is the  equivalent relation of
(\ref{energy}), considering the FRW cosmological model as the
metric inside the Schwarzschild black hole. Thus, the universe of
this type has a quantized mass of the order of the Planck mass
$\rm  M_{pl}=2.18\times 10^{-8}Kg$. These results are similar to
those obtained by other methods
\cite{rosen,louko,kastrup,makela,makela2}.

If one goes over from mass to energy units one finds the following
figure for the ground state energy
\begin{equation}
\rm M_0 c^2 =1.22 \times 10^{28} eV,
\end{equation}
and for the excitation to the next state the energy required will be
\begin{equation}
\rm \left(M_1- M_0 \right) c^2 = 0.9 \times 10^{28} eV.
\end{equation}

On the other hand, we can see that Eq. (\ref{uno}) remains
invariant under the dual symmetries,

\begin{equation}
\rm E_s \rightarrow \frac{c^4}{2 G}R_s, \qquad
R_s\rightarrow \frac{2G}{c^4}E_s,
\label{inv}
\end{equation}
in analogy with the case of magnetic and electric charges
\cite{montonen}.

Let us write the equation (\ref{wave1}) in the following form
\begin{equation}
\rm \frac{d^2\psi}{dx^2} + (\alpha^2_n -x^2) \psi=0,
\end{equation}
where $\alpha_n$ is parameter associated with the energy of the nth
eigenstate
\begin{equation}
\rm \alpha_n^2 = \frac{1}{2} \frac{R_s E_s}{\ell_{pl} E_{pl}}
=2(n+\frac{1}{2}), \qquad thus, \qquad \alpha_n= \frac{E_s}{E_{pl}},
\end{equation}
and the quantum solution is similar to the harmonic oscillator
case
\begin{equation}
\rm \psi_n(x) = \left(\frac{1}{\sqrt{\pi} n! 2^n}
\right)^{\frac{1}{2}} H_n(x) \, e^{-\frac{1}{2}x^2},
\end{equation}
with $\rm H_n(x)$ the Hermite polynomials.

The area parameter $\rm A=4\pi R_s^2$ has corrections depending on
the nth eigenstate (see Eq. (\ref{area})),
\begin{equation}
\rm A= 2\pi \left[ R_s^2 +\left(\frac{2G}{c^4}\right)^2 E_s^2
\right]= 32\pi(n+ \frac{1}{2}) \ell_{pl}^2.
\end{equation}

The area of the event horizon of a black hole can take only
discrete values, such that, the quanta of the area are of the same
order of magnitude as the Planck area. It is easy to check that
this parameter is invariant under the transformation (\ref{inv})
and is similar to Eq. (\ref{area}).

On the other hand, the calculation of the partition function can
be done along the lines of \cite{OSK}, leading to similar results.

\section{Conclusions}
In this paper using the canonical quantization, a time-independent
Schr\"odinger equation for the case of FRW model inside a
Schwarzschild black hole has been obtained.

This system looks like a quantum linear harmonic oscillator, and
using the creation-annihilation representation we find interesting
relations between the quantities $R_s, E_s, E_{pl}$ and
$\ell_{pl}$, (see(\ref{uno},\ref{energyn},\ref{dos})), in terms of
the discrete parameter $n$. With these relations, we obtain the
discrete mass spectrum for this type of Planck scale closed
universe (\ref{rule mass}). A generalization of this procedure to the supersymmetric quantum black hole will be reported elsewhere.

\noindent {\bf Acknowledgments}\\
We thank Drs. H. Rosu and J.Torres for several useful remarks.
This work was partially supported by CONACyT grant 37851, PROMEP
and Gto. University projects. J.J. Rosales was
supported in part by PROMEP under Grant UGTO-PTC-31.

\end{document}